\documentclass[slac_one]{revtex4}
\usepackage{graphicx}
\usepackage{fancyhdr}
\usepackage{psfrag}
\usepackage{braket}
\usepackage{color}
\definecolor{Red}{rgb}{1.,0.,0.}
\newcommand{\Red}[1]{{\color{Red}{#1}}}

\begin{document}

\title{Chiral loop corrections in $B \to D l \nu_l$ decays} 

%
\author{Jan O. Eeg}
\affiliation{University of Oslo, Physics Department, P. O. Box 1048 Blindern, N-0316 Oslo 3, Norway}
\author{Svjetlana Fajfer}
\email[Electronic address:]{svjetlana.fajfer@ijs.si}
\affiliation{J. Stefan Institute, Jamova 39, P. O. Box 3000, 1001 Ljubljana, Slovenia}
\affiliation{Department of Physics, University of Ljubljana, Jadranska 19, 1000 Ljubljana, Slovenia}
\author{Jernej F. Kamenik}
\email[Electronic address:]{jernej.kamenik@ijs.si}
\affiliation{J. Stefan Institute, Jamova 39, P. O. Box 3000, 1001 Ljubljana, Slovenia}
\affiliation{INFN, Laboratori Nazionali di Frascati, I-00044 Frascati, Italy}

\begin{abstract}
We determine chiral loop corrections to the B meson decay amplitudes to positive and negative parity charmed mesons within a framework which combines heavy quark and chiral symmetries. 
We find that corrections due to states of opposite parity are competitive with the contributions arising from K and eta meson loops. 
Since lattice studies rely on the chiral behavior of the amplitudes we discuss the chiral limit of our results. 
We also comment on the extraction of $1/m_B$, $1/m_D$ subleading 
form factors relevant for the studies of $B_q \to D_q \tau \nu_\tau$ decays, which are sensitive to possible helicity-suppressed new physics contributions.

\end{abstract}

\maketitle

\thispagestyle{fancy}


The $b\to c$ transitions have been very important in the extraction of the $V_{cb}$ CKM matrix element. 
In experiments aimed to determine $V_{cb}$, actually the product
$|V_{cb} {\cal F } (1) |$ is extracted, where  ${\cal F } (1)$ denotes
the $B \to D$ or $B \to D^*$ hadronic form factors at zero recoil. A
lack of precise information about the shapes of these form factors
away from zero recoil is one of the main sources of uncertainties.
Furthermore recent spectroscopic discoveries of low lying orbitally excited charmed meson states at the charm and B factories have prompted reevaluation of their contributions to B meson decays. 
The leading order (LO) $SU(3)$ chiral corrections at leading and next to leading order (NLO) in $1/m_Q$ to the $B_q\to D_q^{(*)}$ semileptonic form factors have previously been computed~\cite{Boyd:1995pq}. Also virtual effects of positive parity heavy meson states in these transitions have been considered before~\cite{Falk:1993iu}. 
However, phenomenological discussion at the time was limited by the lack of experimental and lattice  QCD (LQCD) information available on the relevant phenomenological parameters. We are now able to provide the first more reliable estimates of these contributions, but we also extend the program to B meson transitions to positive parity charmed mesons~\cite{EFK1}. In particular we address the issue of extrapolation of LQCD results on these quantities to the chiral regime.
In the $B_q\to D_q \tau\nu$ decay mode due to the large tau mass one can discuss helicity suppressed contributions as such coming from the exchange of charged scalars~\cite{Tanaka:1994ay,Kiers:1997zt,Nierste:2008qe,Kamenik:2008tj}. 
This calls for high precision theoretical estimates for both the dominating vector 
as well as the subleading relative scalar form factor contribution. 
We have estimated the leading chiral symmetry breaking corrections, governing the differences between the vector and scalar form factors in $B_q\to D_q$ transitions~\cite{EFK2}. These corrections can on the one hand be used to guide LQCD studies in their chiral extrapolations. On the other hand we use them to estimate qualitatively the relative scalar form factor values in $B_s\to D_s$ transitions.

First we present the most important results of our calculation~\cite{EFK1} of
leading chiral loop corrections to the form factors governing transitions of B mesons to positive and negative parity charmed mesons. In our heavy meson chiral perturbation theory (HM$\chi$PT) description, heavy-light mesons appear in velocity ($v$) dependent spin-parity doublets due to heavy quark spin symmetry, while chiral $SU(N_f)$ symmetry determines their interactions with the multiplet of light pseudo-Goldstone bosons. Such a framework provides a systematic expansion terms of light and inverse heavy quark masses (for details c.f. ref.~\cite{Fajfer:2006hi}).   
The relevant form factor contributions in the effective theory are obtained using operator bosonization procedure introducing the universal Isgur-Wise (IW) functions $\xi(w)$, $\tilde \xi(w)$ and $\tau_{1/2}(w)$ (where $w=v\cdot v'$ is the product of velocities of initial and final heavy-light states) governing transitions within the negative, positive parity doublets, and between states of opposite parity respectively. We calculate loop corrections to these effective weak vertices coming from the one loop diagram topologies shown in Fig. 1.
\psfrag{pi}[bl]{\footnotesize $\Red{ {\pi^i(q)}}$}
\psfrag{Ha}[cc]{\footnotesize $\Red{H_a(v)}$}
\psfrag{Hb}[cc]{\footnotesize $\Red{H_b(v')}$}
\psfrag{Hc}[cc]{\footnotesize $~\Red{H_c(v)}$}
\psfrag{Hd}[cc]{\footnotesize $~~\Red{H_c(v')}$}
\begin{figure*}[t]
\centering
\includegraphics[width=145mm]{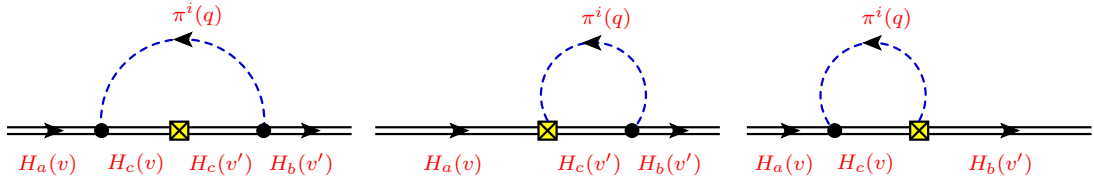}
\caption{Weak vertex correction diagrams. Crossed boxes represent effective weak vertices while filled circles represent LO strong vertices. Only diagrams of the utmost left topology contribute to the amplitude at the leading chiral log order.} \label{JACpic2-f2}
\end{figure*}
In addition, wave function renormalization of the heavy-light fields needs to be taken into account. This has been done e.g. in ref.~\cite{Fajfer:2006hi} and non-zero contributions come from the self energy diagrams with LO effective strong couplings in the loop.
Presence of two opposite parity heavy-light meson multiplets in the theory introduces a new scale -- the mass gap between the ground and excited heavy meson states $\Delta_{SH}$.
In order to tame the chiral behavior of the
amplitudes containing the mass gap we use the $1/\Delta_{SH}$
expansion of the chiral loop integrals~\cite{Fajfer:2006hi}.
We have argued~\cite{Becirevic:2006me} 
that the expansion works well in a $SU(2)$ theory where kaons and etas,
whose masses would compete with $\Delta_{SH}$, do not
propagate in the loops. Therefore we can write down explicit expressions
for the chiral loop corrected IW functions specifically for
the strangeless states in the $SU(2)$ theory\footnote{Following~\cite{EFK1} we absorb the infinite and scale dependent pieces from one loop amplitudes into the appropriate counterterms at order $\mathcal O (m_q)$.}:
\begin{eqnarray}
[\xi(w)]^{\mathrm{Loop}}  &=&   \xi(w) \Bigg\{ 1 + \frac{3}{32\pi^2 f^2} m^2_{\pi} \log \frac{m^2_{\pi}}{\mu^2}  \Bigg[ g^2 2 (r(w)-1) 
 - h^2  \frac{m^2_{\pi}}{4\Delta_{SH}^2} \left(1-w\frac{\tilde \xi(w)}{\xi(w)}\right) - h g \frac{m^2_{\pi}}{\Delta_{SH}^2} w(w-1)\frac{\tau_{1/2}(w)}{\xi(w)}\Bigg] \Bigg\},\nonumber\\
\label{eq:5}
\end{eqnarray}
where
$r(x) = {\log(x+\sqrt{x^2-1})}/{\sqrt{x^2-1}}$ and all IW functions on the rhs should be considered tree-level, similarly
\begin{eqnarray}
[\tau_{1/2 }(w)]^{\mathrm{Loop}} &=&  \tau_{1/2}(w) \Bigg\{ 1 + \frac{3}{32\pi^2 f^2} m^2_{\pi} \log \frac{m^2_{\pi}}{\mu^2}  \Bigg[ - g\tilde g(2r(w)-1) - \frac{3}{2} (g^2+\tilde g^2)\nonumber\\
&&
\hskip -0cm + h^2  \frac{m^2_{\pi}}{4\Delta_{SH}^2} \left(w-1\right) - h g \frac{m^2_{\pi}}{2\Delta_{SH}^2} \frac{\xi(w)}{\tau_{1/2}(w)} w(1+w) + h\tilde g \frac{m^2_{\pi}}{2\Delta_{SH}^2} \frac{\tilde \xi(w)}{\tau_{1/2}(w)} w(1+w) \Bigg] \Bigg\}.
\label{eq:6}
\end{eqnarray}
The first parts of Eqs.~(\ref{eq:5}) and~(\ref{eq:6}) contain the leading contributions while the calculated  $1/\Delta_{SH}$ corrections are contained in the second parts. Definitions and values for the LO strong couplings $g$, $\tilde g$ and $h$ can be found in ref.~\cite{EFK1}.
We present
the chiral behavior of the IW functions
in the chiral limit below the $\Delta_{SH}$ scale in Fig. 2.
\psfrag{xk1}[bc]{{$r\sim m_{u,d}/m_s$}}
\psfrag{xm1}[tc][tc][1][90]{{${\xi'(1)}^{\mathrm{Loop}}/\xi'(1)^{\mathrm{Tree}}$}}
\psfrag{s1}[cl]{\footnotesize{$(1/2)^-$ contributions}}
\psfrag{s3}[cl]{\footnotesize{$\xi'(1)-\tilde \xi'(1)=1$}}
\psfrag{s4}[cl]{\footnotesize{$\xi'(1)-\tilde \xi'(1)=-1$}}
\begin{figure*}[t]
\centering
\includegraphics[width=85mm]{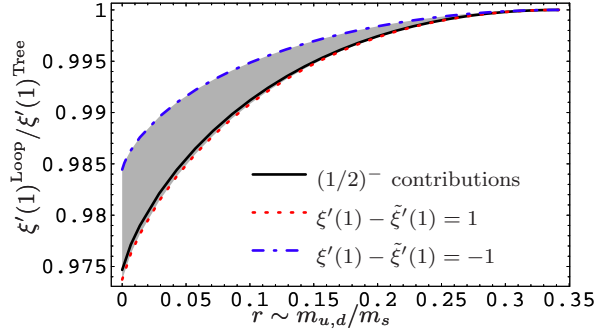}
\caption{Chiral extrapolation of the slope of the IW
function at $w=1$ ($\xi'(1)$). Negative parity heavy states'
contributions (black line) and a range of possible positive parity
heavy states' contribution effects when the difference of slopes of
$\xi(1)$ and $\tilde \xi(1)$ is varied between $1$ (red dashed line)
and $-1$ (blue dash-dotted line)} \label{JACpic2-f3}
\end{figure*}


Next we focus particularly on the matrix element of the vector $b\to c$ quark current between a B and a D meson of velocity $v$ and $v'$ respectively, which can be parametrized in terms of two form factors:
\begin{eqnarray}
&& \bra{D(v')}\bar c \gamma_{\mu} b \ket{B(v)} = \sqrt{m_B m_D} [ h_+(w) (v+v')_{\mu} 
+ h_-(w) (v-v')_{\mu} ]\,.\label{eq:h+expansion}
\end{eqnarray}
It is important to know that in heavy quark expansion 
$h_+(w)$ is the leading order term in $1/m_Q$ (proportional to the IW function $\xi(w)$) while $h_-(w) \sim 1/m_c - 1/m_b$. 
After the bosonization of the relevant new operators which appear at $1/m_Q$~\cite{EFK2}, we determine leading logarithmic chiral corrections relevant for LQCD extraction of the $h_+(w)$ and $h_-(w)$ form factors in the $SU(2)$ theory:
\begin{eqnarray}
\label{eq:hpchi}
	h_+ &=& h_+^{\mathrm{Tree}} \left[ 1 + 3 g^2 \frac{r(w)-1}{(4\pi f_{\pi})^2}  m_{\pi}^2\log \frac{m_{\pi}^2}{\mu^2} + m_{\pi}^2 c_{+}(\mu,w) \right]\,, \\
\label{eq:hmchi}
	h_- &=&  h_-^{\mathrm{Tree}} \left\{ 1 - 3 g^2 \frac{2 - Y^{*}_+ [r(w)+1]}{(4\pi f_{\pi})^2}  m_{\pi}^2\log \frac{m_{\pi}^2}{\mu^2} \right.
+ m_{\pi}^2 c_{-}(\mu,w) \Big\}\,,
\end{eqnarray}
where $Y^{*}_+= h_-^{*\mathrm{Tree}}/h_-^{\mathrm{Tree}}$ is the ratio of vector current matrix elements proportional to $v-v'$ between vector and pseudoscalar states respectively (see ref.~\cite{EFK1} for details and other possible choices of parameterizing this quantity), and the $w$ dependence is implicit. Here and in the rest of the text $c_i(\mu,w)$ denote the sums of local analytic counter-terms, which cancel the $\mu$ dependence of the chiral log pieces. As stressed by the notation, they will in general have a non-trivial $w$ dependence, originating both from the finite analytic residuals of chiral loops as well as from local NLO chiral current operators, needed to cancel the UV divergences of chiral loops. 
The ratio $h_-/h_+$ is particularly important since it enters in the multiplicative contribution to the differential decay rate distribution distinguishing $B\to D \tau\nu$ from $B \to D e\nu$ decays. The chiral corrections in this case read:
\begin{eqnarray}
	\frac{h_-}{h_+} &=&  \left(\frac{h_-}{h_+}\right)^{\mathrm{Tree}} \left[ 1 - 3 g^2 \frac{r(w)+1}{(4\pi f_{\pi})^2}  m_{\pi}^2\log \frac{m_{\pi}^2}{\mu^2}  \right]
+ \left(\frac{h^*_-}{h_+}\right)^{\mathrm{Tree}} 3 g^2 \frac{r(w)+1}{(4\pi f_{\pi})^2}  m_{\pi}^2\log \frac{m_{\pi}^2}{\mu^2} + m_{\pi}^2 c_{{\pm}}(\mu,w)\,,
\end{eqnarray}
as written in ~\cite{EFK2} and where relevant quantities are defined. 

We can summarized our results:
within a HM$\chi$PT  framework, which includes even and odd parity
heavy meson interactions with light pseudoscalars as
pseudo-Goldstone bosons, we have calculated chiral loop corrections
to the functions $\xi$ and $\tau_{1/2}$. As in previous cases~\cite{Fajfer:2006hi,Becirevic:2004uv}  we have  shown that
the leading pionic chiral logarithms are not changed by the
inclusion of even parity heavy meson states we consider chiral
extrapolation of IW functions. Our results are particularly
important for the LQCD extraction of the form factors. The
present errors on the $V_{cb}$ parameter in the exclusive channels
are of the order few percent. This calls for careful control over
theoretical uncertainties in its extraction. 
We have found also~\cite{EFK2} how calculating the ratio of current matrix elements in $B\to D$ and $B^*\to D^*$ transitions on the lattice can improve the extraction of the scalar form factor. 



\begin{acknowledgments}
This work is supported in part by the European Commission RTN network, 
Contract No. MRTN-CT-2006-035482 (FLAVIAnet).
The work of S.F. and J.K. is supported in part by the Slovenian
Research Agency. J.O.E. is supported in part by the Research Council
of Norway. 
\end{acknowledgments}

\end{document}